\documentstyle[12pt]{article}
\begin{document}
\begin{center}
{\large\bf On the
Hamiltonian Description of Fluid Mechanics} \\
\vskip 1cm
{\bf I.Antoniou$^{1,2}$, G.P.Pronko$^{1,3}$}\\

$^{1}${\it International Solvay Institute, Brussels, Belgium}\\

$^{2}${\it Department of Mathematics, Aristoteles University of
Thesaloniki,
Ellas}\\

$^{3}${\it Institute for
High Energy Physics , Protvino, Moscow reg., Russia}
\end{center}

\begin{abstract}
We suggest the Hamiltonian approach for fluid mechanics based on the
dynamics, formulated in terms of Lagrangian variables. The construction of
the canonical variables of the fluid sheds a light on the origin of Clebsh
variabes, introduced in the previous century. The develoted formalism
permitts to relate the circulation conservation (Tompson theorem) with the
invariance of the theory with respect to special diffiomorphisms and
estublish also the new conservation laws. We discuss also the difference
of the Eulerian and Lagrangian description, pointing out the incompeteness
of the first. The constructed formalism is also applicable for ideal
plasma. We conclude with several remarks on the quantization of the fluid.
\end{abstract}

\section{Introduction}

At the present time almost all fundamental physical phenomena could be
formulated  in the frameworks of either classical or quantum mechanics.
That means that these phenomena admits the Hamiltonian description, which
due to its long history developed many powerful methods of analysis of the
general properties of evolution of the systems and the tools for the
solutions
of partial problems. The Hamiltonian formalism also provides the unique
way of transition from classical to quantum description of the systems.

In this respect the fluid mechanics stands aside (in spite of its name)
from orthodox mechanics. The reasons for that is not only the infinite
number of degrees of freedom of fluid. We have already learned how to
formulate the theory of classical and quantum fields. The main difference
between conventional field theory and the fluid is that in the first case
we can speak about the dynamics of the field at one point in space (which
of course interacts with the field at the neighbouring points), while in
the case of the fluid, describing the interaction of the neighbouring
particles which constitute the fluid  we are loosing its position in the
space due to the motion of fluid. In the same time the objective of usual
problems of hydrodynamics
is to define the velocity, density and a thermodynamical variable
(pressure or entropy) as the functions of the coordinates $\vec x$ and
time $t$ for the appropriate boundary conditions and/or initial data
\cite{Landau}. The similar problems also appear for magnetohydrodynamics
dealing with sufficiently dense plasma \cite{Chen}. For the developing of
the Hamiltonian formalism we need to start with more detailed description
based initially on the trajectories of the particles which constitutes the
fluid or plasma. This description is especially important for plasma,
because the fundamental electromagnetic interaction could be formulated
only in terms of the trajectories of the charges.
Needless to say that some aspects of this approach was extensively studied
in the series of papers by J.Marsden, A.Weinstein, P.Kupershmidt, D.Nolm ,
T.Ratiu and C.Levermore \cite{Mars} especially in the context of stability
problem for
ideal fluid. In contrast to these papers here we deliberately avoid, when
it is possible, the use of language of modern differential geometry to
make our paper accessible to physicists.

Of course not all
properties of fluid could be formulated in the frameworks of the
Hamiltonian approach. For example, we leave open the question of the
energy dissipation, viscosity et cetera.

\section{The Lagrangian equations of motion}

In fluid mechanics there are two different pictures of description. The
first, usually refereed as Eulerian, uses as the coordinates  the space
dependent fields of velocity, density and some thermodynamic variable.
The second, Lagrangian description, uses the coordinates of the particles
$\vec x(\xi_i,t)$  labeled by the set of the parameters $\xi_i$, which
could be considered as the initial positions $\vec\xi=\vec x(t=0)$
 and time $t$. These initial positions  $\vec \xi$ as well, as the
coordinates $\vec x(\xi_i,t)$ belong to some domain $D \subseteq R^3$. 
In sequel we shall consider only conservative systems, where the paths of
different particles do not cross, therefore it is clear  that  the
functions $\vec x(\xi_i,t)$ define a diffeomorphism of  $D \subseteq R^3$
and the inverse functions $\vec \xi(x_i,t)$ should also exist.
\begin{eqnarray}\label{1}
x_j(\xi_i,t)\Big|_{\vec \xi=\vec\xi(x_i,t)}&=x_j,\nonumber\\
\xi_j(x_i,t)\Big|_{\vec x=\vec x(\xi_i,t)}&=\xi_j.
\end{eqnarray}
The density of the particles in space at time $t$ is
\begin{equation}\label{2}
\rho(\vec x,t)=\int d^3 \xi \rho_0(\xi_i)\delta(\vec x-\vec x(\xi_i,t)),
\end{equation}
where $\rho_0(\xi)$ is the initial density at time $t=0$.
The velocity field  $\vec v$ as a function of coordinates $\vec x$ and $t$
is:
\begin{equation}\label{3}
\vec v(x_i,t)=\dot{\vec x}(\vec\xi(x_i,t),t),
\end{equation}
where $\vec\xi(x,t)$ is the inverse function (\ref{1}). The velocity also
could be written in the following form:
\begin{equation}\label{4}
\vec v(x_i,t)=\frac{\int d^3 \xi \rho_0(\xi_i)\dot{\vec
x}(\xi_i,t)\delta(\vec
x-\vec x(\xi_i,t))}{\int d^3 \xi \rho_0(\xi_i)\delta(\vec x-\vec
x(\xi_i,t))},
\end{equation}
or
\begin{equation}\label{5} 
\rho (x_i,t)\vec v(x_i,t)=\int d^3 \xi \rho_0(\xi_i)\dot{\vec
x}(\xi_i,t)\delta(\vec x-\vec x(\xi_i,t)).
\end{equation}
Let us calculate the time derivative of the density using its definition
(\ref{2}) :
\begin{eqnarray}\label{6}
&\dot \rho (x_i,t)=\displaystyle\int d^3 \xi
\rho_0(\xi_i)\frac{\partial}{\partial t}
\delta(\vec x-\vec x(\xi_i,t))\nonumber\\
&=\displaystyle\int d^3 \xi \rho_0(\xi_i)\left(-\dot{\vec
x}(\xi_i,t)\right)\frac{\partial}{\partial \vec x}
\delta(\vec x-\vec x(\xi_i,t))\nonumber\\
&=-\frac{\partial}{\partial \vec x}\int d^3 \xi
\rho_0(\xi_i)\dot{\vec x}(\xi_i,t)\delta(\vec x-\vec
x(\xi_i,t))\nonumber\\
&=-\displaystyle\frac{\partial}{\partial \vec x}\rho (x_i,t)
\vec v (x_i,t)
\end{eqnarray}
In such a way we verify the continuity equation of fluid dynamics:
\begin{equation}\label{7}
\dot \rho (x_i,t)+\vec\partial\Bigl(\rho (x_i,t)\vec v (x_i,t)\Bigr)=0.
\end{equation}.

Using the coordinates $\vec x(\xi_i,t)$ as a configurational variables we
can consider the simplest motion of the fluid described by the Lagrangian
\begin{equation}\label{8}
L=\int d^3 \xi\frac{m\dot{\vec x}^2 (\xi_i,t)}{2}.
\end{equation}
The equations of motion which follow from (\ref{8}) apparently are
\begin{equation}\label{9}
m\ddot{\vec x} (\xi_i,t)=0
\end{equation}
Now let us find what does this equation mean for the density and velocity
of the fluid. For that we shall differentiate both sides of (\ref{5}) with
respect to time
\begin{eqnarray}\label{10} 
&\displaystyle\frac{\partial}{\partial t}\rho (x_i,t)\vec v(x_i,t)=\int
d^3 \xi \rho_0(\xi_i)\ddot{\vec x}(\xi_i,t)\delta(\vec x-\vec
x(\xi_i,t))\nonumber\\
&+\int d^3 \xi
\rho_0(\xi_i)\dot{\vec x}(\xi_i,t)\displaystyle\frac{\partial}{\partial
t}\delta(\vec x-\vec x(\xi_i,t))
\end{eqnarray}
The first term in the r.h.s. of (\ref{10}) vanishes due to the equations
of motion(\ref{9}) and transforming the second in the same way, as we did
in (\ref{6}) we arrive at
\begin{equation}\label{11}
\displaystyle\frac{\partial}{\partial t} \rho (x_i,t)\vec v(x_i,t)+
\displaystyle\frac{\partial}{\partial x_k}\Bigl(\rho (x_i,t)
\vec v (x_i,t) v_k(x_i,t)\Bigr)=0
\end{equation}
Let us rewrite (\ref{11}) in the following form:
\begin{eqnarray}\label{12} 
&\vec v(x_i,t)\Big[\dot \rho (x_i,t)+\displaystyle\frac{\partial}{\partial
x_k}\Bigl(\rho (x_i,t) v_k (x_i,t)\Bigr)\Big]\nonumber\\
&+\rho(x_i,t)\Bigl[\dot{\vec
v}(x_i,t)+v_k(x_i,t)\displaystyle\frac{\partial}{\partial
x_k} \vec v (x_i,t)\Bigr]=0.
\end{eqnarray}
The first term in (\ref{12}) vanishes due to the continuity equation,
while the second gives Euler's equation in the case of the free flow:
\begin{equation}\label{13} 
\dot{\vec v}(x_i,t)+
v_k(x_i,t)\displaystyle\frac{\partial}{\partial
x_k} \vec v (x_i,t)=0
\end{equation}

To move further we need to introduce into the Lagrangian (\ref{8}) the
"potential energy" term which will give rise to the  internal
pressure field in Euler's equation. As we have mentioned above, the
functions $\vec x(\xi_i,t)$ define a diffeomorphism,in $R^3$ therefore the
Jacobean matrix
\begin{equation}\label{14}
A^j_k(\xi_i,t)=\frac{\partial x_j(\xi_i,t)}{\partial \xi_k}
\end{equation}
is nondegenerate for any $\xi_i$ and $t$. The integral (\ref{1}) may be
expressed via the Jacobean determinant of $A^j_k(\xi_i,t)$:
\begin{equation}\label{15}
\rho(\vec x,t)=\frac{\rho_0(\xi_i)}{detA(\xi_i,t)}\Bigg|_{\vec \xi=\vec
u(x_i,t)}.
\end{equation}
Assume for the simplicity that the initial density $\rho_0(\xi_i)$ is
uniform in $D\subseteq R^3$, and  effectively normalize the density field
$\rho(\xi_i,t)$ by putting $\rho_0(\xi_i)=1$ (one particle in the
elementary volume). Then (\ref{15}) becomes
\begin{equation}\label{16}
\rho(\vec x,t)=\frac{1}{detA(\xi_i,t)}\Bigg|_{\vec \xi=\vec
u(x_i,t)}.
\end{equation}
Consider as the "potential energy" the  functional of $detA$:
\begin{equation}\label{17}
L=\int d^3 \xi\Biggl[\frac{m\dot{\vec x}^2
(\xi_i,t)}{2}-f(detA(\xi_i,t))\Biggr].
\end{equation}
Now the equations of motion become
\begin{equation}\label{18}
m\ddot{x_j} (\xi_i,t)-\displaystyle\frac{\partial}{\partial
\xi_k}\Biggl(\left(A^{-1}\right)_j\,^k(\xi_i,t) f'(detA)detA\Biggr)=0
\end{equation}
Substituting $\ddot{x_j} (\xi_i,t)$ from (\ref{18}) to the
equation(\ref{10}) and acting as we did in the derivation of the equation
(\ref{13}), we obtain
\begin{equation}\label{19}
m\rho (x_i,t)\displaystyle\left(\frac{\partial}{\partial t}+v_k(x_i,t)
\frac{\partial}{\partial
x_k}\right)v_j(x_i,t)-\displaystyle\frac{\partial}{\partial
x_j}\left(f'(detA(\xi_i,t))\Bigg|_{\vec \xi=\vec
u(x_i,t)}\right)=0
\end{equation}

It is now obvious that if we identify the
$-1/mf'(detA(\xi_i,t))\Big|_{\vec \xi=\vec
u(x_i,t)}$ with pressure $p(x_i,t)$, the equation (\ref{19}) takes the
form of usual Euler equation without viscosity:
\begin{equation}\label{20}
\rho (x_i,t)\displaystyle\left(\frac{\partial}{\partial t}+v_k(x_i,t)
\frac{\partial}{\partial
x_k}\right)v_j(x_i,t)=-\displaystyle\frac{\partial}{\partial
x_j}p(x_i,t).
\end{equation}
Note, that the pressure $p(x_i,t)$ which appeared here is not due to an
external force, but the result of interaction between particles. The
further complication of the Lagrangian  is not necessary for the moment.
One comment nevertheless should be done. No modification of the Lagrangian
will give us the Navier-Stokes equation because the later includes the
dissipation effects, which is not time symmetric....

\section{Hamiltonian formalism}

The  Hamiltonian formalism of classical mechanics,
which among other adventures gives us the unique way for construction of
the quantum theory. In this respect the hydrodynamics stands aside because
its fundamental variables - local velocity, density and thermodynamical
function does not admit an immediate Hamiltonian interpretation, in spite
of quite often in the text-book one can meet the terms like "momentum
density" (see e.g. \cite{Landau}). 

We shall introduce the canonical variables according to the Lagrangian
(\ref{17}), so our canonical coordinates will be the functions $\vec
x(\xi_i,t)$ and its conjugated momenta are defined as the derivatives of
the Lagrangian with respect to the velocities $\dot {\vec x}(\xi_i,t)$:
\begin{equation}\label{1.1} 
\vec p(\xi_i,t)=\frac{\delta L}{\delta \dot {\vec x}(\xi_i,t)}= m\dot
{\vec x}(\xi_i,t),
\end{equation}
The Hamiltonian is given by the Legendre transform of
the Lagrangian:
\begin{equation}\label{1.2}
H=\int d^3 \xi \left(\frac{1}{2m}\vec
p^2(\xi_i,t)+f(detA(\xi_i,t))\right). 
\end{equation}
The canonical Poisson brackets is defined by 
\begin{equation}\label{1.3}
\{p_j(\xi_i), x_k(\xi'_i)\}=\delta_{jk}\delta^3(\xi_i-\xi'_i)
\end{equation}
Apparently the Poisson brackets (\ref{1.3}) and the Hamiltonian
(\ref{1.2}) define the  equations of motion for the canonical variables,
which are equivalent to the Lagrange equations. The phase space $\Gamma$
of the fluid is formed by $\vec x(\xi_i,t),\vec p(\xi_i,t)$. 
What is missing at the moment is the space interpretation of the variables
$\vec x(\xi_i,)$ and $\vec p(\xi_i)$. To get one let us introduce the new
objects using the same averaging, as we used in the previous section
\begin{equation}\label{1.4}
\vec l(x)=\int d^3 \xi\vec p(\xi_i)\delta(\vec x-\vec x(\xi_i))=
\rho(x)\vec p(\xi_i(x)).
\end{equation}
This new object will serve as a part of the coordinates in the phase space
$\Gamma$, therefore we need to calculate its Poisson brackets, using
(\ref{1.3}). The result has the following form:
\begin{equation}\label{1.5}
\{l_j(x_i),l_k(y_i)\}=\Bigl[\l_k(x_i)\frac{\partial}{\partial
x_j}+\l_j(y_i)\frac{\partial}{\partial x_k}\Bigr]\delta(\vec x-\vec y). 
\end{equation}
The Poisson brackets was introduced geometrically as "the
hydrodynamic-type
brackets" long ago in the papers  \cite {DubrovinNovikov} without
physical derivation. The present discussion reveals the origin of
these brackets. The commutation relation (\ref{1.5}) coincides with
algebra of 3-dimensional diffeomorphisms, where the $l_j(x_i)$ serves as
the generators. In other words, with the help of these generators we can
realize the finite  diffeomorphism $x_j\rightarrow \phi_j(x_i)$ of any
$x$-dependent dynamical variable. It should be mentioned that the group of
diffeomorphisms, generated by  $l_j(x_i)$ is not a gauge symmetry in case
of fluid mechanics, as it is in the case of e.g.
relativistic string or membrane. In the same time in fluid mechanics there
is an infinite dimensional symmetry with respect to special (i.e. volume
preserving) diffeomorphisms $SDiff$, which will be considered in the next
section.

As we mentioned above, the $l_j(x_i)$ is only a part of $x$-dependent
coordinates in the phase space $ \Gamma $, one may consider it as the 
"$x$-dependent momenta". Now we should define the corresponding 
"$x$-dependent coordinates". For example we may choose for this role the
functions $\xi_j(x_i)$, which are the inverse to $x_j(\xi_i)$ functions
(\ref{1}). Differentiating the first equation (\ref{1}) with respect to
$x$ we obtain:
\begin{equation}\label{1.6}
A^j_k(\xi_i(x_k))\frac{\partial \xi_k(x_i)}{\partial x_l}=\delta ^j_l,
\end{equation}
in other words, the matrix
\begin{equation}\label{1.7}
a^k_l(x_i)=\frac{\partial \xi_k(x_i)}{\partial x_l}
\end{equation}
is inverse to $A^j_k(\xi_i(x_k))$. Therefore , from (\ref{16}) follows
that
\begin{equation}\label{1.8}
\rho(x_i)=det a(x_i)
\end{equation}
To simplify the calculation of the Poisson brackets we can express
$\xi_j(x_i)$ in the following form:
\begin{equation}\label{1.9}
\xi_j(x_i)=\frac{\int d^3 \xi \xi_j\delta(\vec x-\vec x(\xi_i,t))}{\int
d^3 \xi \delta(\vec x-\vec x(\xi_i,t))}.
\end{equation}
From (\ref{1.9}) we easily obtain
\begin{equation}\label{1.10}
\{\xi_j(x_i),\xi_k(y_i)\}=0
\end{equation}
The calculation of the Poisson brackets between $\xi_j(x_i)$ and
$l_j(x_i)$
is more involved, but the result is simple:
\begin{equation}\label{1.11}
\{l_j(x_i),\xi_k(y_i)\}=-\frac{\partial \xi_k(x_i)}{\partial
x_j}\delta(\vec x-\vec y)
\end{equation}
In such a way we have constructed the set of $x$-dependent coordinates
$\left(l_j(x_i), \xi_j(x_i)\right)$ in the phase space $\Gamma$, but the
transformation
\begin{equation}\label{1.12}
\left(p_j(\xi_i),x_j(\xi_i)\right) \rightarrow \left( l_j(x_i),
\xi_j(x_i)\right)
\end{equation}
is not  canonical.  The set of canonical $x$-dependent coordinates in
$\Gamma$ could be obtained in the following way. Let us multiply both
sides of (\ref{1.11}) by matrix $A^j_m(\xi_i(x_k))$:
\begin{equation}\label{1.13}
A^j_m(\xi_i(x_k))\{l_j(x_i),\xi_k(y_i)\}=
-\delta ^k_m\delta(\vec x-\vec y),
\end{equation}
where we have used (\ref{1.6}). Further, due to  the relation
(\ref{1.10}), we can put $A^j_m(\xi_i(x_k))$ inside the brackets and
obtain:
\begin{equation}\label{1.14}
\{\pi_m(x_i),\xi_k(y_i)\}=\delta ^k_m\delta(\vec x-\vec y),
\end{equation}
where 
\begin{equation}\label{1.15}
\pi_m(x_i)=-A^j_m(\xi_i(x_k))l_j(x_i)
\end{equation}
By direct calculation we also obtain
\begin{equation}\label{1.16}
\{\pi_m(x_i),\pi_k(y_i)\}=0,
\end{equation}
so the set $\left(\pi_m(x_i),\xi_k(y_i)\right)$ is formed by the canonical
variables. In terms of these canonical variables the generators of the
group of diffiomorphisms $l_j(x_i)$ has the following form:
\begin{equation}\label{1.17}
l_j(x_i)=-\frac{\partial \xi_k(x_i)}{\partial x_j}\pi_k(x_i).
\end{equation}
For the Lagrangian considered, from (\ref{1.1}) and (\ref{1.4}) follows
that $l_j(x_i)$ is given by
\begin{equation}\label{1.18}
l_j(x_i)=m\rho(x_i)v_j(x_i)
\end{equation}
and the representation (\ref{1.17}) is very similar to Clebsh
parametrization  \cite{Clebsh} of the velocity. The distinction of
(\ref{1.17}) from the original Clebsh paramerization is the appearance of
three "potentials", instead of two, for the 3-dimensional fluid. The
reason of this difference will be discussed in the end of this section.

The above given construction of the canonical $x$-dependent variables is
the most simple one. For some problems another set might be  useful. Let
us construct one denoting by $\zeta_j(x_i)$ the canonical coordinates:  
\begin{equation}\label{1.19}
\zeta_j(x_i)=\int d^3 \xi \xi_j\delta(\vec x-\vec x(\xi_i,t)).
\end{equation}
Calculating the Poisson brackets of $\zeta_j(x_i)$ and $l_j(x_i)$ we
obtain:
\begin{equation}\label{1.20}
\{l_j(x_i),\zeta_k(y_i)\}=\zeta_k(x_i)\frac{\partial}{\partial x_j}\delta
(\vec x-\vec y).
\end{equation}
From (\ref{1.20}) we conclude that $l_j(x_i)$ should have the following
representation
\begin{equation}\label{1.21}
l_j(x_i)=\zeta_k(x_i)\frac{\partial}{\partial x_j}\eta_k(x_i),
\end{equation}
where $\eta_k(x_i)$ is the variable, canonically conjugated to
$\zeta_j(x_i)$:
\begin{equation}\label{1.22}
\{\eta_j(x_i),\zeta_k(y_i)\}=\delta_{jk}\delta (\vec x-\vec y).
\end{equation}

The canonical Hamiltonian (\ref{1.2}) should now be expressed in the terms
of new variables $\left(\pi_m(x_i),\xi_k(y_i)\right)$. Indeed, let us
insert  the unity
\begin{equation}\label{1.23}
1=\int d^3 x\delta(\vec x-\vec x(\xi_i))
\end{equation}
into the integrand (\ref{1.2}) and change the order of integration:
\begin{equation}\label{1.24}
H=\int d^3 x\int d^3 \xi \left(\frac{1}{2m}\vec p^2(\xi_i,t)+f(det
A(\xi_i))\right)\delta(\vec x-\vec x(\xi_i)), 
\end{equation}
Fulfilling the integration over $\xi$ with the help of (\ref{2}) and
(\ref{1.4}) we obtain:
\begin{equation}\label{1.25}
H=\int d^3 x \left(\frac{1}{2m\rho (x_i)}\vec l^2(x_i)+\rho
(x_i)f(\frac{1}{\rho(x_i)}))\right). 
\end{equation}
Making use of (\ref{1.17}) we can express $H$ via canonical variables
$\left(\pi_m(x_i),\xi_k(y_i)\right)$:
\begin{equation}\label{1.26}
H=\int d^3 x \left(\frac{1}{2m\rho (x_i)}\frac{\partial
\xi_k(x_i)}{\partial
x_j}\frac{\partial \xi_m(x_i)}{\partial x_j}\pi_k(x_i)\pi_m(x_i)+V(\rho
(x_i))\right),
\end{equation}
where we have introduced a notation $V(\rho (x_i))$ for the "potential"
part of the energy. This term represents  the internal energy of the fluid
and it should vanish (up to inessential constant) for homogeneous density
distribution $\rho (x_i)=\rho_0$. A phenomenological expression for
$V(\rho (x_i))$ could be written as follows \cite{Zakharov}:
\begin{equation}\label{1.27}
V(\rho (x_i))=\frac{\kappa}{2\rho_0} (\delta \rho (x_i))^2 +\lambda
(\nabla \rho (x_i))^2+... \quad,
\end{equation}
where $\delta \rho (x_i)$ is the deviation of the density from its
homogeneous distribution:
\begin{equation}\label{1.28}
\delta \rho (x_i)=\rho (x_i)-\rho_0.
\end{equation}
The first term in (\ref{1.27}) is responsible for the  sound wave in the
fluid ($\kappa$ is the velocity of sound), the second term in (\ref{1.27})
describes the  dispersion of the sound waves.

In order to reveal the relation of the Hamiltonian flow, generated by
(\ref{1.26}) with geodesic flow \cite{Arnold} let us introduce the metric
tensor $g_{jk}(x_i)$:
\begin{equation}\label{1.29}
g_{jk}(x_i)=A^m_j(\xi_i(x_k))A^m_k(\xi_i(x_k)).
\end{equation}
With this notation (\ref{1.26}) takes the form:
\begin{equation}\label{1.30}
H=\int d^3 x
\left(\frac{1}{2m}\sqrt{g(x_i)}g^{km}(x_i)\pi_k(x_i)\pi_m(x_i)+
V(\frac {1}{\sqrt{g(x_i)}}) \right).
\end{equation}
The metric tensor with upper indices $g^{jk}(x_i)$ denotes, as usually
the inverse matrix and 
\begin{equation}\label{1.31}
g(x_i)=det g_{jk}(x_i)=\frac{1}{\rho ^2(x_i)}
\end{equation}
The representation (\ref{1.30}) permit us to consider the hydrodynamics as
the geodesic flow on the dynamical manifold with  metric $g_{jk}(x_i)$ and
many general properties of the hydrodynamics could be derived from this
fact (see e.g. \cite{Arnold}).

We shall complete this section with one important comment concerning our
description of hydrodynamics. In the present discussion the phase space of
the 3-dimensional fluid is 6-dimensional, as it naturally follows from our
approach. This could be compared with recent paper \cite{Jackiw}, where
(in our notations) only $\rho(x_i)$ and $v_j(x_i)$ are regarded as the
coordinates in the phase space (this point of view of  could be found also
in different text books and articles) see for example \cite{Arnold}, \cite
{Zakharov}, \cite {Lanczos}.

According to the conventional point of view the state of the fluid is 
determined by its velocity and density and therefore all other variables
like ours $\vec x(\xi_i)$ are not needed.
Indeed, solving the Euler equations of motion we can express the velocity
$\vec v(x_i,t)$ and the density $\rho (x_i,t)$ at time $t$ via the initial
data $\vec v_(x_i,0)$ and $\rho_ (x_i,0)$ . Then, from the definition of
$\vec v(x_i,t)$ (\ref{3}) follows
\begin{equation}\label{1.32}
\vec v\left(x(\xi,t),t\right))=\dot {\vec x}(\xi,t).
\end{equation}
(here we have suppressed the indices of the arguments for brevity).
Apparently we can solve these equations with respect to $\vec x(\xi,t)$,
provided we know the initial data $\vec x(\xi,0)$. Therefore it may seem
that the variables $\vec x(\xi,t)$ are unnecessary, as it could be
obtained through the others. But the initial data $\vec x(\xi,0)$ are
half of the  canonical variables in Hamiltonian formalism. So, in our
approach we indeed need more variables. These additional variables provide
{\it the complete description} of the fluid in a sense that solving the
equations of motion we define not only the $\vec v(x_i,t)$ and $\rho
(x_i,t)$ but also the trajectories of the particles which could not be
obtained in the conventional formalism. The situation is analogous to the
rigid body rotation: here the  phase space $\Gamma$ is formed by 3 angles
and 3 components of the angular momentum $J_i$. As the Hamiltonian depends
on the components of $J_i$ only, we can consider separately the evolution
of the angular momentum. This is incomplete description for this
mechanical system. The complete one certainly should includes the
evolution of the angles, which define the location of the rigid body in
space.

In the case of  fluid dynamics {\it the complete description} is to be
done in the 6-dimensional phase space $\Gamma$ formed by $\vec x(\xi)$ and
$\vec p(\xi)$ (or $\vec \xi(x)$ and $\vec l(x)$). If, as it is usually the
case, the Hamiltonian depends only on $\vec l(x)$ and $\rho(x)$, partial
description in terms of the velocities and densities considered as
"relevant" variables is possible. This means that we do not care about the
evolution of  the coordinates of the phase space which are considered as
inessential. The "relevant" part of the coordinates does not necessarily
form a simplectic subspace in $\Gamma$.  The nondegenerate Poisson
brackets in $\Gamma$ could become degenerate on the subset of $\Gamma$,
corresponding to the "relevant" variables. This is indeed the case in the
rigid body and in the conventional fluid dynamics. In the first case the
degeneracy of the algebra of Poisson brackets for the "relevant" variables
--- angular momentum is well known. Its center element is  Casimir
operator of the rotation group. In the case of fluid dynamics the algebra
of the "relevant" variables --- velocities and densities has the following
form:
\begin{eqnarray}\label{1.33}
&\{v_j(x_i),v_k(y_i)\}&=-\frac{1}{\rho(x)}\left(\nabla_jv_k(x_i)-
\nabla_kv_j(x_i)\right)\delta(\vec x-\vec y)\nonumber\\
&\{v_j(x_i),\rho(y_i)\}&=\nabla_j\delta(\vec x-\vec y)\nonumber\\
&\{\rho(x_i),\rho(y_i)\}&=0
\end{eqnarray} 
This algebra could be obtained from equations (\ref{1.5}), (\ref{1.10})
and (\ref{1.11}). The center of this algebra is infinite-dimensional and
its structure  depends on the dimension of $x$-space. In
2-dimensional case the center is formed by :
\begin{equation}\label{1.34}
I_k=\int dxdy \left(\frac{\partial v_1(x,y)}{\partial y}-
\frac{\partial v_2(x,y)}{\partial x}\right)^k \rho^{1-k}(x,y),
\end{equation}
see \cite{Arnold}, \cite{Jackiw} for the discussion.
In 3-dimentional case the general answer is not known to our knowledge and
its construction is out of the scope of our paper. As an example we
mention that Hopf invariant 
\begin{equation}\label{1.35}
Q=\int d^3 x \epsilon_{jkl}v_j(\vec x)\nabla_k v_l(\vec x)
\end{equation}
belongs to this center.

\section{Infinite-dimensional symmetry and integrals of motion.}

As we have mentioned above, the Lagrangian (\ref{17}) possesses the
invariance with respect to the "volume preserving" group of
diffeomorphisms $SDiff[D]$, where $D\subseteq R^3$. Indeed, let us write
the Lagrangian (\ref{17}) in terms of the Lagrangian density
\begin{equation}\label{2.1}
L=\int d^3 \xi {\cal L}(\xi_i)
\end{equation}
and consider the transformations from $SDiff[D]$ of the coordinates
$\xi_i\in D$ 
\begin{equation}\label{2.2}
\xi_j\rightarrow \xi'_j=\phi_j(\xi_i),
\end{equation}

\begin{equation}\label{2.3}
det\frac{\partial \phi_j(\xi_i)}{\partial \xi_k}=1.
\end{equation}
Apparently, due to (\ref{2.3}) we obtain:
\begin{equation}\label{2.4}
L=\int d^3 \phi(\xi_i) {\cal L}\left(\phi(\xi_i)\right)=\int d^3 \xi {\cal
L}\left(\phi(\xi_i)\right)
\end{equation}
and according to Noether's theorem this invariance results in the
existence of an infinite set of integrals of motion. To obtain these
integrals we first need to find the parametrization of the transformations
(\ref{2.2}), (\ref{2.3}) in the vicinity of identity transformation:
\begin{equation}\label{2.5}
\phi_j (\xi_i)=\xi_j+\alpha_j(\xi_i).
\end{equation}
From (\ref{2.3}) follows the equation for $\alpha_j(\xi_i)$:
\begin{equation}\label{2.6}
\frac{\partial \alpha_j(\xi_i)}{\partial \xi_j}=0.
\end{equation}
Further we must explicitly take into account that the volume preserving
diffeomorphism (\ref{2.2}) leaves the boundary of $D$ invariant. We
shall limit ourself with the case when $D$ is formed by extraction of the
domain with the differentiable boundary given by  
\begin{equation}\label{2.7}
g(\xi_i)=0
\end{equation}
from $R^3$. Physically that means that we put in the fluid the fixed body,
the shape of which  is given by (\ref{2.7}). The condition that the
infinitesimal
diffeomorphism (\ref{2.5}) preserves $D$ in this case is
\begin{equation}\label{2.8}
g(\xi_j+\alpha_j(\xi_i))\Big|_{g(\xi_i)=0}=0,
\end{equation}
or
\begin{equation}\label{2.9}
\alpha_j(\xi_i)\nabla_j g(\xi_i)\Big|_{g(\xi_i)=0}=0.
\end{equation}
Geometrically equation (\ref{2.9}) means that the vector $\vec
\alpha(\xi_i)$ is tangent to the surface, defined by (\ref{2.7}),
because the vector $\nabla_j g(\xi_i)\Big|_{g(\xi_i)=0}$ is proportional
to the normal $n_j(\xi_i)$ of the surface (\ref{2.7}) at the point
$\xi_i$.

From Noether's theorem we obtain that the invariance of the Lagrangian
with respect to the transformation (\ref{2.5}) gives the following
conservation law:
\begin{equation}\label{2.10}
\frac{\partial}{\partial t} \int_{D} d^3 \xi p_m(\xi_i)\frac{\partial
x_m(\xi_i)}{\partial \xi_l} \alpha_l(\xi_i)=0,
\end{equation}
where $\alpha_l(\xi_i)$ satisfies to the conditions (\ref{2.6}) and
(\ref{2.9}). The existence of these conditions forbids to take the
variation
of the l.h.s. of (\ref{2.10}) over $\alpha_l(\xi_i)$ and obtain the local
form
of integrals of motion. For that we need to extract from (\ref{2.6}) and
(\ref{2.9}) the integral properties of $\alpha_l(\xi_i)$. 

Consider an arbitrary, single-valued,  differentiable in $D$ function
$\beta(\xi_i)$. Then the following equations are valid:
\begin{eqnarray}\label{2.11}
&0=\displaystyle\int_{D}d^3 \xi \beta(\xi_i)\frac{\partial
\alpha_j(\xi_i)}{\partial
\xi_j}=\nonumber\\
&\displaystyle=\int_{D}d^3 \xi \frac{\partial}{\partial
\xi_j} \Bigl(\beta(\xi_i)\alpha_j(\xi_i)\Bigr)-\int_{D}d^3 \xi
\alpha_j(\xi_i)\frac{\partial\beta(\xi_i)}{\partial\xi_j},
\end{eqnarray}
The first equality is valid due to condition (\ref{2.6}). Using
Stokes theorem we can transform the integral of total derivative in the
last equality (\ref{2.11}):
\begin{equation}\label{2.12}
\displaystyle\int_{D}d^3 \xi \frac{\partial}{\partial
\xi_j} \Bigl(\beta(\xi_i)\alpha_j(\xi_i)\Bigr)=\int_{\partial D}
dS_j\Bigl(\beta(\xi_i)\alpha_j(\xi_i)\Bigr)=0,
\end{equation}
where $\partial D$ denotes the boundary of $D$. The last integral in
(\ref{2.12}) vanishes due to  condition (\ref{2.9}) because the
differential $dS_j$ is proportional to the normal vector of the surface
$\partial D$, defined by (\ref{2.7}). From (\ref{2.11}) and (\ref{2.12})
we conclude that
\begin{equation}\label{2.13}
\int_{D}d^3 \xi
\alpha_j(\xi_i)\frac{\partial\beta(\xi_i)}{\partial\xi_j}=0
\end{equation}
for any smooth, differentiable $\beta(\xi_i)$. Taking this property of
$\alpha_j(\xi_i)$ into account we obtain from the conservation laws
(\ref{2.10}) that the quantities
\begin{equation}\label{2.14}
J_k(\xi_i)=p_m(\xi_i)\frac{\partial x_m(\xi_i)}{\partial \xi_k} 
\end{equation}
are conserved modulo some term which is the gradient of a scalar. In
particular, that means that
\begin{equation}\label{2.15}
R_j(\xi_i)=\epsilon_{jkl}\frac{\partial}{\partial\xi_k}J_l(\xi_i)=
\epsilon_{jkl}\frac{\partial}{\partial
\xi_k}\left(p_m(\xi_i)\frac{\partial x_m(\xi_i)}{\partial \xi_l}\right). 
\end{equation}
is the integrals of motion.
Note, that as the group of invariance is infinite-dimensional, we obtain
an infinite number of integrals of motion. With respect to Poisson
brackets
(\ref{1.3}) the $R_j(\xi_i)$'s form an algebra. This algebra could be
written in a compact form for integrated objects
\begin{equation}\label{2.16}
R[\phi]=\int d^3 \xi \phi_j(\xi_i)R_j(\xi_i),
\end{equation}
where $\phi_j (\xi_i)$ are smooth, rapidly decreasing functions. The
algebra of $R[\phi]$ induced by Poisson brackets (\ref{1.3}) has the
form:
\begin{equation}\label{2.17}
\{R[\phi],R[\psi]\}=R[curl\phi\times curl\psi]
\end{equation}
The construction of the $x$-dependent object, corresponding to
$R_j(\xi_i)$ is not an easy task, because our "averaging" with $\delta
(\vec x-\vec x(\xi_i))$ will introduce time dependence and instead of
conserved object we shall obtain a density, whose time derivative gives a
divergence of a "current". Therefore we need to introduce another kind of
averaging without explicit refering to the $\vec x(\xi_i)$ coordinates.
For that recall that under diffeomorphism a closed loop transforms into
closed loop. Then let us consider such a loop $\lambda$ and a surface
$\sigma$ whose boundary is $\lambda$. The integral
\begin{equation}\label{2.18}
V=\int_{\sigma} dS_j R_j(\xi_i)
\end{equation}
where the vector $dS_j$ is as usually the area element times the vector,
perpendicular to the surface, is conserved, because of the conservation of
$R_j(\xi_i)$. Further, from Stokes theorem we have:
\begin{equation}\label{2.19}
V=\oint_{\lambda} d \xi_j p_m(\xi_i)\frac{\partial x_m(\xi_i)}{\partial
\xi_j}.
\end{equation}
Changing the variables in (\ref{2.19}) we obtain:
\begin{equation}\label{2.20}
V=\oint_{\Lambda}d x_j \frac{l_j(x_i)}{\rho (x_i)}=\oint_{\Lambda}d x_j
v_j(x_i),
\end{equation}
where $\Lambda$ is the image of the loop $\lambda$ under diffeomorphism
$\xi_j\rightarrow x_j(\xi_i)$. The object (\ref{2.20}) is very well known
in hydrodynamics as the "circulation" and its conservation is known as
W.Thompson theorem \cite{Thompson}. The relation of the circulation
conservation with the invariance under special diffeomorphisms  was first
explicitly established in \cite{Salmon}, though it also could be extracted
from general discussion in Appendix 2 of \cite{Arnold}.

Conservation of circulation is not the only consequence of (\ref{2.10}). 
Consider for example the case of 2-dimensional space. Here, instead of the
conserved vector $R_j(\xi_i)$ we shall have the conserved scalar
\begin{equation}\label{2.21}
R(\xi_i)=\epsilon_{kl}\frac{\partial}{\partial\xi_k}J_l(\xi_i)=
\epsilon_{kl}\frac{\partial}{\partial
\xi_k}\left(p_m(\xi_i)\frac{\partial x_m(\xi_i)}{\partial \xi_l}\right), 
\end{equation}
This scalar defines the following integrals of motion:
\begin{equation}\label{2.22}
I_n=\int_D d^2 \xi R^n(\xi_i)=\int_D
d^2\xi\Biggl(\epsilon_{kl}\frac{\partial}{\partial
\xi_k}\left(p_m(\xi_i)\frac{\partial x_m(\xi_i)}{\partial
\xi_l}\right)\Biggr)^n.
\end{equation}
Changing variables in (\ref{2.22}) $\xi_j\rightarrow \xi_j(x_i)$ which is
possible, because $\xi_j(x_i)$ is a diffeomorphism of $D$, we obtain:
\begin{equation}\label{2.23}
I_n=\int_D d^2 x \rho(x_i)\Biggl(\epsilon_{kl}A^j_k(\xi(x))A^m_l(\xi(x))
\frac{\partial p_m(\xi(x))}{\partial x_j}\Biggr)^n,
\end{equation}
where the matrix $A^j_k(\xi)$ was defined in (\ref{14}).
We can present $I_n$ as
\begin{equation}\label{2.25}
I_n=\int_D d^2 x \rho(x_i)^{1-n}\Biggl(\epsilon_{jm}
\frac{\partial}{\partial x_j}\frac{l_m(x_i)}{\rho(x_i)}\Biggr)^n.
\end{equation}
using the following
property of 2-dimensional matrix $A^j_k(\xi)$:
\begin{equation}\label{2.24}
\epsilon_{kl}A^j_k(\xi(x))A^m_l(\xi(x))=\epsilon_{jm}det A(\xi(x))=
\epsilon_{jm}\frac{1}{\rho(x_i)},
\end{equation}

Recall that in the case of the Lagrangian (\ref{17}), we are
considering
\begin{equation}\label{2.26}
\frac{l_j(x_i)}{\rho(x_i)}=mv_j(x_i)
\end{equation}
Therefore the integrals $I_n$ coincide with the center (\ref{1.34}) of
the Poisson algebra (\ref{1.33}).

In the case of 3-dimensional space we can construct the analogous
integrals of motion, integrating the products of the vector
(\ref{2.15}):
\begin{equation}\label{2.27}
K_{j_1,j_2...j_n}=\int_D d^3 \xi
R_{j_1}(\xi_i)R_{j_2}(\xi_i)...R_{j_n}(\xi_i)
\end{equation}
Changing variables  $\xi_j\rightarrow \xi_j(x_i)$ as above we shall
obtain:
\begin{equation}\label{2.28}
R_j(\xi)\rightarrow R_j(\xi(x))=
\epsilon_{jkl}A^m_k(\xi(x))A^n_l(\xi(x))\frac{\partial
p_n(\xi(x))}{\partial x_m}.
\end{equation}
In the 3-dimensional case the matrices $A^m_j(\xi(x))$ satisfy the
equation:
\begin{equation}\label{2.29}
\epsilon_{jkl}A^m_k(\xi(x))A^n_l(\xi(x))=\epsilon_{mnr}\frac{\partial\xi_j
(x)}{\partial x_r}det A(\xi(x)),
\end{equation}
Therefore (\ref{2.28}) takes the form:
\begin{equation}\label{2.30}
R_j(\xi(x))=
\frac{1}{\rho(x_i)}\epsilon_{mnr}\frac{\partial\xi_j(x)}{\partial
x_r}\frac{\partial p_n(\xi(x))}{\partial x_m}
\end{equation}
The integrals $K_n$ in (\ref{2.27}) become
\begin{equation}\label{2.31}
K_{j_1,j_2...j_n}=\int_{D} d^3 x
\rho(x_i)R_{j_1}(\xi(x))R_{j_2}(\xi(x))...R_{j_n}(\xi(x))
\end{equation}
Apparently, due to the presence of  $\;\frac{\partial\xi_j(x)}{\partial
x_r}\;$ in (\ref{2.30}), we can not in this case  express (\ref{2.31})
only in terms of velocity and density, i.e. the Eulerian description does
not admit  this kind of integrals of motion.

In the 3-dimensional case there is one more integral, which does not 
exist in any other dimension. Recall that the vector $J_k(\xi_i)$, given
by (\ref{2.14}) is conserved modulo gradient, therefore the integral
\begin{equation}\label{2.32}
Q=\int_{D}d^3 \xi J_k(\xi_i)R_k(\xi_i)
\end{equation}
is conserved because $\vec R(\xi_i)=curl\vec J(\xi_i)$. 
Transforming the  $\xi_j$-dependent variables to the $x_j$-dependent ones
in (\ref{2.32}) we obtain Hopf invariant:
\begin{equation}\label{2.33}
Q=\int_{D}d^ x \epsilon_{jkl}p_j(\xi(x))\frac{\partial
p_k(\xi(x))}{\partial x_l}.
\end{equation}

\section{Inclusion of the electromagnetic interaction. Plasma.}

We shall consider plasma as a fluid of two components, namely electrons
with mass $m$ and electric charge $(-e)$ and ions with mass $M$
and charge $(+e)$. The coordinates of electrons we shall denote as $\vec
x(\xi_i)$, while the coordinates of ions will be $\vec X(\Xi_i)$.  The
interaction of the components of the plasma   with the electromagnetic
field $A_{\mu}(x)$ is governed by the following Lagrangian:
\begin{eqnarray}\label{3.1} 
L=&L_0^{el}+L_0^{ion}+\displaystyle
\int d^3 \xi \Biggl[e\left(A_0(\vec x(\xi_i,t))-\dot {\vec x}(\xi_i,t)\vec
A(\vec x(\xi_i,t))\right)\nonumber\\
&-e\left(A_0(\vec x(\xi_i,t))-\dot {\vec X}(\Xi_i,t)\vec A(\vec
X(\Xi_i,t))\right)\Biggr]\nonumber\\
&-\frac{1}{4}\displaystyle \int d^3 x F^{\mu\nu}(x,t)F_{\mu\nu}(x,t),
\end{eqnarray}
where $L_0^{el}$ and $L_0^{ion}$ are "free" Lagrangians
\begin{eqnarray}\label{3.2} 
&L_0^{el}=\displaystyle\int d^3 \xi \Biggl[\frac{m\dot{\vec
x}^2(\xi_i,t)}{2}
- f^{el}(det\frac{\partial x_j(\xi_i,t)}{\partial
\xi_k})\Biggr],\nonumber\\
&L_0^{ion}=\displaystyle\int d^3 \Xi\Biggl[\frac{M\dot{\vec
X}^2(\Xi_i,t)}{2}-f^{ion}(det\frac{\partial X_j(\Xi_i,t)}{\partial
\Xi_k})\Biggr],
\end{eqnarray}
and $F_{\mu\nu}(x,t)$ denotes the electromagnetic field tensor:
\begin{equation}\label{3.3}
F_{\mu\nu}(x,t)=\partial _{\mu}A_{\nu}(x,t)-\partial _{\nu}A_{\mu}(x,t)
\end{equation}
The advantage of Lagrangian description of fluid (plasma) is clear.
Using of the coordinates of charged particles as the fundamental variables
makes explicit the interaction with electromagnetic field.

The Lagrangian (\ref{3.1}) possesses  $U(1)$ gauge invariance. We shall
use the usual Hamiltonian formalism for the constraint system
\cite{Dirac}. The canonical variables are:
\begin{eqnarray}\label{3.4}
\vec x(\xi_i), &\qquad & \vec p(\xi_i)=\frac{\delta L}{\delta \dot{\vec
x}(\xi_i)};\nonumber\\
\vec X(\Xi_i), &\qquad  &\vec P(\Xi_i)=\frac{\delta L}{\delta \dot{\vec
X}(\Xi_i)};\nonumber\\
\vec A(x_i),&\qquad & \vec P_{em}(x_i)=\frac{\delta L}{\delta \dot{\vec
A}(x_i)}=-\vec E(x_i);\nonumber\\
A_0(x_i),&\qquad & P_{em}^0(x_i)=0,
\end{eqnarray}
where $\vec E(x_i)$ is the electric field strength:
\begin{equation}\label{3.5}
\vec E(x_i)=-\nabla A_0(x_i)-\dot{\vec A}(x_i).
\end{equation}
The last of the equations (\ref{3.4}) is actually the primary constraint.
The Legendre transformation of the Lagrangian (\ref{3.1}) gives us the
canonical Hamiltonian:
\begin{eqnarray}\label{3.6}
&H=\int d^3 x \Bigl[\frac{1}{2}\left(\vec P_{em}^2 (x_i)+\vec
H^2(x_i)\right)+A_0(x_i)\nabla \vec \Pi(x_i)\Bigr]+\nonumber\\
&\int d^3 \xi \Bigl[\frac{1}{2m}\left(\vec p(\xi_i)+e\vec
A(x(\xi_i))\right)^2-eA_0(x(\xi_i))+f^{el}(det\frac{\partial
x_j(\xi_i,t)}{\partial \xi_k})\Bigr]+ \nonumber\\
&\int d^3 \Xi \Bigl[\frac{1}{2M}\left(\vec P(\Xi_i)-e\vec
A(X(\Xi_i))\right)^2+eA_0(X(\Xi_i))+f^{ion}(det\frac{\partial
X_j(\Xi_i,t)}{\partial \Xi_k})\Bigr],
\end{eqnarray}
where $\vec H(x_i)=curl\vec A(x_i)$ is the magnetic field strength.The
requirement of the conservation of the primary constraint $\Pi_0(x_i)=0$
gives the secondary constraint ( Gauss law ):
\begin{equation}\label{3.7} 
\nabla_j
\left(P_{em}(x_i)\right)_j-e\left(\rho_{el}(x_i)-\rho_{ion}(x_i)\right)=0.
\end{equation}
Now we can add to the primary constraint $\Pi_0(x_i)=0$ the gauge fixing
condition $A_0(x_i)=0$ and eliminate these variables from consideration.
Introducing the $x$-dependent functions instead of $u$ and $U$-dependent,
as we did in the 2-nd section we obtain the Hamiltonian of the plasma in
the following form:
\begin{eqnarray}\label{3.8}
&H=\int d^3 x \Bigl[\frac{1}{2}\left(\vec P_{em}^2 (x_i)+\vec
H^2(x_i)\right)
+\displaystyle\frac{\left(\vec l(x_i)+e\rho_{el}(x_i)\vec
A(x_i)\right)^2}{2m\rho_{el}(x_i)}\nonumber\\
&+\displaystyle\frac{\left(\vec L(x_i)+e\rho_{ion}(x_i)\vec
A(x_i)\right)^2}{2m\rho_{ion}(x_i)}+v_{el}(\rho_{el}(x_i))+v_{ion}(\rho_{i
on}(x_i))\Bigr],
\end{eqnarray}
where we have introduced
\begin{eqnarray}\label{3.9}
\vec l(x_i)&=\int d^3 \xi \vec p(\xi_i)\delta (\vec x-\vec
x(\xi_i))\nonumber\\
\vec L(x_i)&=\int d^3 \Xi \vec P(\Xi_i)\delta (\vec x-\vec X(\Xi_i))
\end{eqnarray}
This Hamiltonian is gauge invariant with respect to the transformations,
generated by the constraint (\ref{3.7}). Further we impose the Coulomb
gauge condition on the electromagnetic field
\begin{equation}\label{3.10}
\nabla_j A_j (x_i)=0
\end{equation}
and following the usual procedure will eliminate the longitudinal
components of $\vec A(x_i)$ and $\vec P_{em}(x_i)$. As a result of the
gauge fixing, the longitudinal part of $\vec P_{em}(x_i)$ give rise to the
Coulomb term in the Hamiltonian 
\begin{equation}\label{3.11}
H_{Col}=-e^2\int d^3 x\left(\rho_{el}(x_i)-\rho_{ion}(x_i)\right)
\left(-\frac{1}{4\pi|\vec x-\vec y|}\right)
\left(\rho_{el}(y_i)-\rho_{ion}(y_i)\right).
\end{equation}
and the whole Hamiltonian takes the following form:
\begin{eqnarray}\label{3.12}
&H=H_{Col}+\int d^3 x \Bigl[\frac{1}{2}\left(\vec P_{em\bot}^2 (x_i)+\vec
H^2(x_i)\right)\nonumber\\
&+\displaystyle\frac{\left(\vec l(x_i)+e\rho_{el}(x_i)\vec
A_\bot(x_i)\right)^2}{2m\rho_{el}(x_i)}+v_{el}(\rho_{el}(x_i))\nonumber\\
&+\displaystyle\frac{\left(\vec L(x_i)+e\rho_{ion}(x_i)\vec
A_\bot(x_i)\right)^2}{2m\rho_{ion}(x_i)}+v_{ion}(\rho_{ion}(x_i))\Bigr],
\end{eqnarray}
where the subscript $\bot$ denotes the transverse components of
electromagnetic variables, for which the Poisson (Dirac) brackets are
given by \cite{Dirac}
\begin{equation}\label{PB}
\{P^{em}_{\bot j}(x),A_{\bot k}(y)\}=\left(\delta_{jk}-\frac{1}{\triangle}
\partial_j\partial_k\right)\delta(\vec x-\vec y). 
\end{equation}
As usually, the Coulomb term (\ref3.11) contains an infinite additive part
of "self interaction"  which should be subtracted. In equilibrium
plasma the Coulomb interaction is known to be screened by the cloud and
the residual Debye interaction is a short rang one (see e.g.
\cite{Balescu}). In \cite{FaddeevNiemi}
it was stated that Coulomb term is reduced to the local functional of the
difference of charge densities
$\left(\rho_{el}(x_i)-\rho_{ion}(x_i)\right)$.

Following our consideration of a fluid in the Section 3, we can introduce
the canonical coordinates $\vec{\xi}(x_i), \vec{\pi}(x_i)$ for electron
and $\vec{\Xi}(x_i), \vec{\Pi}(x_i)$ for ion components of the plasma:
\begin{eqnarray}\label{3.13}
\{\pi_j(x_i),\xi_k(y_i)\}=\delta_{jk}\delta(\vec x-\vec y), \nonumber\\
\{\Pi_j(x_i),\Xi_k(y_i)\}=\delta_{jk}\delta(\vec x-\vec y),
\end{eqnarray}
for which 
\begin{eqnarray}\label{3.14} 
&l_j(x_i)=\displaystyle\vec{\xi}(x_i)\frac{\partial
\vec{\pi}(x_i)}{\partial x_j},\nonumber\\
&L_j(x_i)=\displaystyle\vec{\Xi}(x_i)\frac{\partial
\vec{\Pi}(x_i)}{\partial x_j}.
\end{eqnarray}
Substituting (\ref{3.14}) into (\ref{3.12}) we obtain the Hamiltonian in
terms of the canonical variables. 

The Lagrangian (\ref{3.1}) is invarian with respect to the volume
preserving diffeomorphisms of both components of plasma separately, so we
shall have in this case two sets of conservation laws -- one for
electrons, the other for ions. Applying the procedure, which we have
described in the previous section we shall construct the conserved
circulations:
\begin{eqnarray}\label{3.15}
&V^{el}_{\Lambda}=\oint_{\Lambda}d x_j \frac{l_j(x_i)}{\rho_{el}
(x_i)},\nonumber\\
&V^{ion}_{\Lambda}=\oint_{\Lambda}d x_j \frac{L_j(x_i)}{\rho_{ion} (x_i)},
\end{eqnarray}
where $l_j(x)$ and $L_j(x)$ are given by (\ref{3.14}). In the same way we
can construct the analogues of the integrals (\ref{2.31}) and (\ref{2.33})
for this case. Note that in the case of plasma the equation (\ref{1.18})
is not valid due to the presence of the electromagnetic field, instead we
have:
\begin{eqnarray}\label{3.16}
&{\vec l}(x)=\rho_{el}(x)\left(m{\vec v}_{el}(x)-e{\vec
A}(x)\right),\nonumber\\
&{\vec L}(x)=\rho_{ion}(x)\left(M{\vec v}_{ion}(x)+e{\vec A}(x)\right),
\end{eqnarray}
therefore the equations (\ref{3.15}) are not the circulations of the
velocities. In the same time, adding circulations $V^{el}_{\Lambda}$ and
$V^{ion}_{\Lambda}$ on the common  contour $\Lambda$ we obtain:
\begin{equation}\label{3.17}
V_{\Lambda}=V^{el}_{\Lambda}+V^{ion}_{\Lambda}=\oint_{\Lambda}d x_j
\left(mv_{el}(x)+Mv_{ion}(x)\right),
\end{equation}
so this linear combination of circulations of electons and ions velocities
is conserved even in the presence of the electromagnatic interaction.

Summarizing we can say that the phase space of plasma $\Gamma$  in Coulomb
gauge is the space with coordinates $\vec A_{\bot}(x), \vec
P^{em}_{\bot}(x); \vec{\xi}(x_i), \vec{\pi}(x_i); \vec{\Xi}(x_i),
\vec{\Pi}(x_i)$ with Poisson brackets given by (\ref{PB}) and
(\ref{3.13}). The evolution of a state in $\Gamma$ is difined by the
Hamiltonian (\ref{3.12}).

\section{Concluding remarks}

The Hamiltonian approach for ideal fluid and plasma, considered in the
present paper in some aspects resembles the Hamiltonian formulation of a
classical field. It deals with the system with infinite number of degrees
of freedom -- "particles" which constitutes a continuous media. These
particles interact with nearest neighboors due to the potential part of
the Lagrangian or Hamiltonian, as the field amplitudes in conventional
field theory. The difference with the later, which we emphasized in the
course of the paper is that the constituents of the fluid change their
position in space while evolution and to have a local description of fluid
in terms of $x$-dependent variables we need the projection realized by
e.g. equation (\ref{4}). The fluid dynamics could be also compared with
the mechanics of the extended objects like $n$-branes. These objects are
also $n$-dimensional media embeded in the space-time, but its constituents
are indistinguishable and due to this property the general diffiomorphisms
of its coordinates appear as the gauge invariance of the theory. In the
case of the fluid it is supposed that the constituents are distinguishable
and, as a result only special diffiomorphisms preserve  the Lagrangian
giving rise to the Noether's integrals of motion.

So far we have considered only classical mechanics of the fluid and
plasma, but as we know any classical Hamiltonian system could be
quantized. Here as a quantization we mean a formal procedure of the
transition from from classical canonical coordinates to the operators,
satisfying commutation relations inherited from Poisson brackets and the
construction of a representation of these operators in an appropriate
space. Of course this procedure has no physical meaning in the case of
plasma, but for a fluid it could be quite reasonable providing us with the
theory of an ideal quantum fluid. The quantum fluid could be considered as
the new representation of a quantum field in which the eigenvalue of the
operator of  density of particles may continuos function of $x$. The
matter is that in the case of usual quantum field we know only Fock
representation \cite{Schweber}, for which the operator of density of
particles in space (or in momentum space) has the eigenvalues which are
the superposition of $\delta$-functions, that means that the average
density is zero and to consider the states with finite density we need
to go outside the Fock space.  

\noindent {\bf Acknowledgements} We would like to thank professor
K.Gustafson and professor A.Razumov for fruitful discussions. The support
of professor I.Prigogine and Solvay Institute is gratefully acknowledged
The work of G.P. was supported in part by Russian Science Foundation Grant
01-01-00201.

\end{document}